\documentclass[sigconf]{acmart}

\usepackage{booktabs}
\usepackage{graphicx}
\usepackage{calc}

\newcommand{\glmm}[7]{$\beta$=#1, SE=#2, $\beta_{std}$=#3, 95\% CI=[#4, #5], z=#6, p#7}

\AtBeginDocument{%
  \providecommand\BibTeX{{%
    \normalfont B\kern-0.5em{\scshape i\kern-0.25em b}\kern-0.8em\TeX}}}

\copyrightyear{2021} 
\acmYear{2021} 
\setcopyright{acmlicensed}\acmConference[CHI '21]{CHI Conference on Human Factors in Computing Systems}{May 8--13, 2021}{Yokohama, Japan} 
\acmBooktitle{CHI Conference on Human Factors in Computing Systems (CHI '21), May 8--13, 2021, Yokohama, Japan}
\acmPrice{15.00}
\acmDOI{10.1145/3411764.3445536}
\acmISBN{978-1-4503-8096-6/21/05}

\begin{document}

\title{Eliciting and Analysing Users' Envisioned Dialogues with Perfect Voice Assistants}

\author{Sarah Theres V\"olkel}
\email{sarah.voelkel@ifi.lmu.de}
\affiliation{%
  \institution{LMU Munich}
  \city{Munich, Germany}
}

\author{Daniel Buschek}
\email{daniel.buschek@uni-bayreuth.de}
\affiliation{%
  \institution{Research Group HCI + AI, Department of Computer Science, University of Bayreuth}
  \city{Bayreuth, Germany}
}

\author{Malin Eiband}
\email{malin.eiband@ifi.lmu.de}
\affiliation{%
  \institution{LMU Munich}
  \city{Munich, Germany}
}

\author{Benjamin R. Cowan}
\email{benjamin.cowan@ucd.ie}
\affiliation{%
  \institution{University College Dublin}
  \city{Dublin, Ireland}
}

\author{Heinrich Hussmann}
\email{hussmann@ifi.lmu.de}
\affiliation{%
  \institution{LMU Munich}
  \city{Munich, Germany}
}

\renewcommand{\shortauthors}{V\"olkel et al.}

\begin{abstract}
  We present a dialogue elicitation study to assess how users envision conversations with a perfect voice assistant (VA). 
In an online survey, N=205 participants were prompted with everyday scenarios, and wrote the lines of both user and VA in dialogues that they imagined as perfect. 
We analysed the dialogues with text analytics and qualitative analysis, including number of words and turns, social aspects of conversation, implied VA capabilities, and the influence of user personality.
The majority envisioned dialogues with a VA that is interactive and not purely functional; it is smart, proactive, and has knowledge about the user. Attitudes diverged regarding the assistant's role as well as it expressing humour and opinions. An exploratory analysis suggested a relationship with personality for these aspects, but correlations were low overall. 
We discuss implications for research and design of future VAs, underlining the vision of enabling conversational UIs, rather than single command ``Q\&As''. 

\end{abstract}



\maketitle

\section{Introduction}
Voice assistants are ubiquitous. They are conversational agents available through a number of devices such as smartphones, computers, and smart speakers~\cite{clark2019state, porcheron2018}, and are widely used in a number of contexts such as domestic~\cite{porcheron2018} and automotive settings~\cite{braun2019}. A recent analysis of more than 250,000 command logs of users interacting with smart speakers~\cite{ammari2019} showed that, whilst people use them for \textit{functional} requests (e.g., playing music, switching the light on/off), they also request more \textit{social} forms of interaction with assistants (e.g., asking to tell a joke or a good night story). Recent reports on smart speakers~\cite{voicereport2020b} and in-car assistant usage trends~\cite{voicereport2020} corroborate these findings, emphasising that voice assistants are more than just speech-enabled ``remote controls''. 

Moreover, voice assistants are perceived as particularly appealing if adapted to user preferences, behaviour, and background~\cite{cowan2015, cowan2016, dahlbaeck2007, lee2003}. Since conversational agents tend to be seen as social actors in general~\cite{nass_1984}, with users often assigning them personalities~\cite{reeves1996}, \textit{personality} has been highlighted as a promising direction for designing and adapting voice assistants.
For example, Braun et al.~\cite{braun2019} found that users trusted and liked a personalised in-car voice assistant more than the default version, especially if its personality matched their own. Although efforts have been made to generate and adapt voice interface personality~\cite{mairesse2010towards}, commercially available voice assistants have so far taken a one-size-fits-all approach, ignoring the potential benefits that adaptation to user preferences may bring.

Systematically adapting a voice assistant to the user is challenging: People tend to show individual differences in preferences for conversations when asked about their envisioned version of a perfect voice assistant~\cite{voelkel2020b}. Personalisation also harbours certain dangers, as an incorrectly matched voice assistant may be less accepted by a user than a default~\cite{braun2019}. Current techniques for generating personalised agent dialogues tend to take a top-down approach~\cite{lee2003, braun2019}, with little user engagement. That is, different versions of voice assistants are developed and then contrasted in an evaluation, without investigating how they should behave in specific tasks or contexts. 

To overcome these problems, we present a pragmatic bottom-up approach, eliciting what users envision are \textit{dialogues with perfect voice assistants}: Concretely, in an online survey, we asked N=205 participants to write what they imagined to be a conversation between a perfect voice assistant and a user for a set of common use cases. In an exploratory approach, we then analysed participants' resulting dialogues qualitatively and quantitatively: We examined the share of speech and interaction between the interlocutors, as well as social aspects of the dialogue, voice assistant and user behaviour, and knowledge attributed to the voice assistant. We also assessed relationships of user personality and conversation characteristics. Specifically we address these research questions:

\begin{enumerate}
    \item RQ1: \textit{How do users envision a dialogue between a user and a perfect voice assistant, and how does this vision vary?}
    \item RQ2: \textit{How does the user personality influence the envisioned conversation between a user and a perfect voice assistant?}
\end{enumerate}

Our contribution is twofold: First, on a conceptual level, we propose a new approach so as to engage users in voice assistant design. Specifically, it allows designers to gain insight into what users feel is the design of a dialogue with a perfect voice assistant. Second, we present a set of qualitative and quantitative analyses and insights into users' preferences for personalised voice assistant dialogues.
This provides much needed information to researchers and practitioners on how to design voice assistant dialogues and how these vary and might thus be adapted to users.

\section{Related Work}
Below we summarise work on the characteristics of today's human-agent conversation, human and conversational agent personality, and adapting the agent to the user.

\subsection{Human-Agent Conversation}
Despite the promise that the name \textit{Conversational Agent} implies, several studies show that conversations with voice assistants are highly constrained, falling short of users' expectations~\cite{reeves_2019, porcheron2018, luger2016, cowan2017}. Existing dialogues with voice assistants are heavily task oriented, taking the form of adjacency pairs that revolve around requesting and confirming actions~\cite{gilmartin2017, clark2019}. Voice assistant research is increasingly interested in imbuing speech systems with abilities that encompass the wider nature of human conversational capabilities, in an attempt to mimic more closely the types of conversations between humans. The abilities to generate social talk~\cite{spillane_2017}, humour~\cite{clark2019}, as well as fillers and disfluencies~\cite{szekely2017synthesising} are being developed as ways of making interaction with speech systems seem more natural. On the other hand, there is scepticism around the benefits this type of naturalness may produce. Recent research on the perception of humanness in voice user interfaces suggests that users tend to perceive voice assistants as impersonal, unemotional, and inauthentic, especially when producing opinions~\cite{doyle2019}. 

Users also tend to see a clear difference between humans and machines as capable interlocutors rather than blurring the boundaries between these two types of partner~\cite{doyle2019}. Machine dialogue partners are regularly seen as ``basic''~\cite{branigan_role_2011} or ``at risk listeners''~\cite{oviatt_linguistic_1998}. To compensate for this, users develop strategies to adapt their speech in interaction~\cite{clark2019state, pelikan2016}. For example, people's speech becomes more formal and precise, with fewer disfluencies and filled pauses along with more command-like and keyword-based structure~\cite{hauptmann1988, oviatt1995, oviatt1996, oviatt2000, lee2009, luger2016}. Moreover, users are also prone to mimic the syntax~\cite{cowan2015} and lexical choices~\cite{branigan2010, branigan_role_2011} of voice assistant's language, a phenomenon termed alignment. This alignment also occurs frequently in human-human interaction but within human-machine dialogue is thought to be driven by a user's attempt to ensure communication success with the system~\cite{branigan2010}. This phenomenon can also be leveraged in conversational design with recent work indicating that users ascribe high likability and integrity to a voice user interface that aligns its language to the user~\cite{linnemann2018}. 
In contrast to prior work, which examined user conversation with voice assistants given the current technological status quo, we take a different approach: We let users freely imagine a conversation they consider to be perfect, using their desired conversation style, syntax, and wording, thus exploring what users actually \textit{do} want when given the choice.

\subsection{Human and Conversational Agent Personality}

Personality describes consistent and characteristic patterns which determine how an individual behaves, feels, and thinks~\cite{mccrae2008}. 
The \textit{Big Five} (also \textit{Five-Factor model} or \textit{OCEAN}) is the most prevalent paradigm for modelling human personality in scientific research and has five broad dimensions~\cite{deRaad2000, deyoung2014, diener1992, goldberg1981, jackson2010, jensen2001, matthews2003, mccrae2008, crae1992, mcniel2006}:

\textit{Openness} reflects an individual's inclination to seek new experiences, imagination, artistic interests, creativity, intellectual curiosity, and an open-minded value and norm system.

\textit{Conscientiousness} reflects a tendency to be disciplined, orderly, dutiful, competent, ambitious, and cautious. 

\textit{Extraversion} reflects a tendency to be friendly, sociable, assertive, dynamic, adventurous, and cheerful.

\textit{Agreeableness} reflects a tendency to be trustful, genuine, helpful, modest, obliging, and cooperative. 

\textit{Neuroticism} reflects an individual's emotional stability and relates to experiencing anxiety, negative affect, stress, and depression. 

A plethora of work in psychology and linguistics has examined the role of personality in human language use~\cite{scherer1979, campbell1978, carment1965, patterson1966, rutter1972, dewaele2000, oberlander2004, pennebaker1999, furnham1990, mehl2006}. This relationship is most pronounced for \textit{Extraversion}. For example, extraverts tend to talk more, use a more explicit and concrete speech style, a simpler sentence structure, and a limited vocabulary with highly frequent words in contrast to introverts
~\cite{furnham1990, oberlander2004, beukeboom2013, dewaele2000, pennebaker1999}. 

Although not common in commercially available voice assistants yet~\cite{vlahos2019}, the construct of personality, in particular the Big Five, has also been leveraged to describe differences in how conversational agents express behaviour~\cite{vinciarelli2014, schmitz2007, trouvain2006, nass2005wired}.
Focusing on voice assistant personality modelling rather than voice assistant personality design, work by~\citet{voelkel2020} points out that the way users describe voice assistant personality may not fit the Big Five model, proposing ten alternative dimensions for modelling a conversational agent personality. Their dimensions such as ``Social-entertaining'' can be expected to be realised by designers also via dialogue-level characteristics, such as including humorous remarks, as we examine here.

\subsection{Adapting the Voice Assistant to the User}
\label{sec:rw:adapt}
Previous work has noted that users enjoy interacting with voice assistants, imbuing them with human-like personality~\cite{cowan2017}. Deliberately manipulating this personality has an impact on user interaction, influencing acceptance and engagement~\cite{cafaro2016, zhou2019}.

Much like human-human interaction, users have preferences for particular personality types, tending to prefer voice assistants who share similar personalities to them~\cite{nass2001, bickmore2005, ehrenbrink2017}, termed the \textit{similarity attraction effect}~\cite{byrne1971, nass2001}. When interacting with a book buying website, extraverted participants showed more positive attitudes towards more extraverted voice user interfaces~\cite{nass2001}, whilst matching a voice user interface's personality to the user's personality also increases feelings of social presence~\cite{lee2003, heeter1992}. Similarity attraction effects have also been seen in in-car voice assistants, whereby users liked and trusted the assistant more if their personalities were matched~\cite{braun2019}. A user's personality also influences their preference for the type of talk voice assistants engage in, with extraverted users preferring a virtual real estate agent that engaged in social talk, and with more introverted users preferring a purely task-oriented dialogue~\cite{bickmore2001, bickmore2005}. While previous work focused on similarity attraction for extraversion in voice assistants, we look at all voice assistant/user personality dimensions.

\subsection{Research Gap}
Overall, related work has provided insights into current shortcomings in voice assistant interaction~\cite{luger2016, cowan2017, porcheron2018} and how users perceive conversations with voice assistants~\cite{clark2019} and their humanness~\cite{doyle2019} contrary to human-human interaction. In contrast to this recent qualitative work, we take a mixed methods approach to explore what users themselves would prefer in a voice assistant dialogue given no technical limitations.

This motivates us to ask users to write their envisioned dialogues with a perfect voice assistant. In this way, we engage users to inform future assistant design, beyond contributing ``compensation strategies'' for current technical limitations. 
Moreover, given the literature's focus on user personality as basis for agent adaptation, we explore relationships of personality and such envisioned dialogues.
Finally, regarding the level of analysis, our study provides the first in-depth dialogue-level assessment, beyond, for example, phrasing of single commands~\cite{braun2019}, social vs functional talk~\cite{bickmore2005}, or nonverbal~\cite{lee2003} investigations of agent personalisation.

\section{Research Design}

\begin{figure}[t!]
    \centering
        \includegraphics[width=\columnwidth]{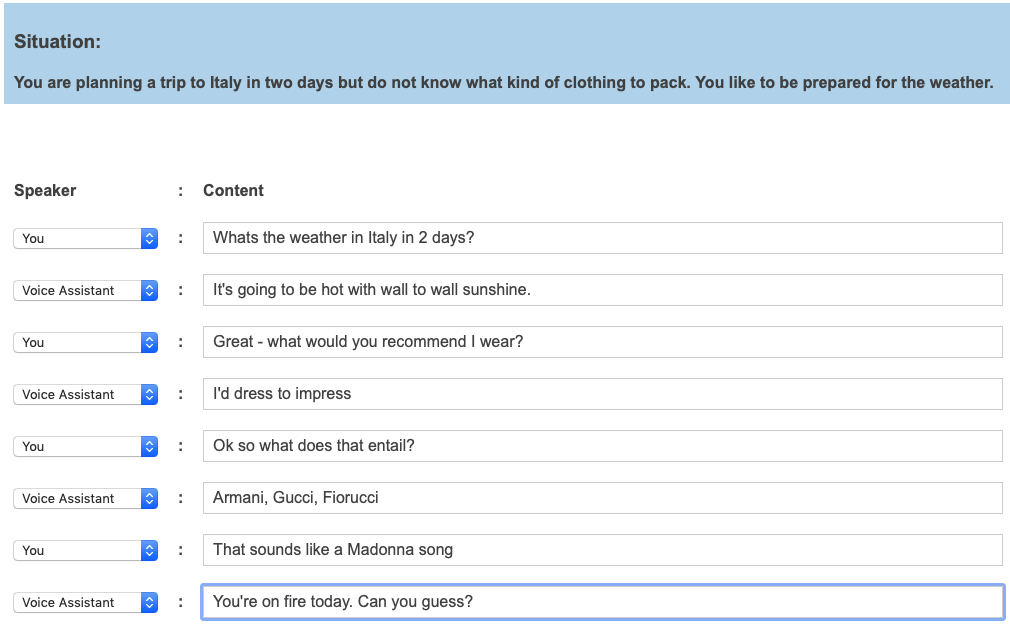}
    \caption{Participants were asked to sketch an envisioned conversation with a perfect voice assistant. For eight given scenarios, they first selected who is speaking from a dropdown menu and then wrote down what the selected speaker is saying. Example dialogue written by participant 28.}
    \label{fig:example_dialogue}
    \Description[Screenshot of the survey task user interfaces]{Screenshot of the survey task user interfaces, showing a set of text entry fields for the dialogue lines: Participants were asked to envision a conversation with a perfect voice assistant. For eight given scenarios, they first selected who is speaking from a dropdown menu and then wrote down what the selected speaker is saying.}
\end{figure}

We conducted an online study to investigate our research questions.
Our research design is inspired by our previous method~\cite{voelkel2020b} which presented participants with different social as well as functional scenarios. In each scenario, participants were asked to complete a dialogue between a user and a voice assistant where the user part was given, that is, they had to add the part of the voice assistant only. We found that there are differences between participants with regard to how they ``designed'' the voice assistant.
These differences were more notable in social scenarios than in functional ones. Moreover, dialogues in functional scenarios were very similar to the current state-of-the-art in interaction with voice assistants. 
    
In our study, we built on this approach, but decided to let participants write entire dialogues (i.e. both the user and the voice assistant part) because we assumed that differences between participants might then emerge more clearly. Participants were presented with different smart home scenarios in which a user solves a specific issue by conversing with the voice assistant (cf. below). We instructed them to write down their envisioned conversation with a ``perfect'' voice assistant, assuming there were no technical limitations to its capabilities, with it being fully capable of participating and engaging in a natural conversation to whatever extent they prefer. Following the Oxford English Dictionary (OED), we define a perfect voice assistant as users' vision of ``complete excellence'' that is ``free from any imperfection or defect of quality''~\cite{OED2020}. Furthermore, we asked participants to ``imagine living in a smart home with a voice assistant''. Hence, we expect they described their version of a conversation in a context of long-term use. This method combines aspects of the story completion~\cite{oro2017} method (i.e. participants writing envisioned interactions) and elicitation approaches~\cite{Villarreal-Narvaez2020} (i.e. asking people to come up with input for a presented outcome), shedding light on user preferences on a technology in the making. 

\begin{table*}[]
\resizebox{\textwidth}{!}{%
\begin{tabular}{@{}ll@{}}
\toprule
\multicolumn{1}{c}{\textbf{Name}} & \multicolumn{1}{c}{\textbf{Description \& Issue}} \\ \midrule
Search & You want to go to the cinema to see a film, but you do not know the film times for your local cinema. \\
Music & You are cooking dinner. You are on your own and you like to listen to some music while cooking. \\
Internet of Things & You are going to bed. You like to read a book before going to sleep. You often fall asleep with the lights on. \\
Volume & You are listening to loud music, but your neighbours are sensitive to noise.\\
Weather & You are planning a trip to Italy in two days but do not know what kind of clothing to pack. You like to be prepared for the weather. \\
Joke & You and your friends are hanging out. You like to entertain your friends, but the group seems to have run out of funny stories. \\
Conversational & You are going to bed, but you are having trouble falling asleep. \\
Alarm &  You are going to bed. You have an important meeting early next morning, and you tend to oversleep.\\
Open Scenario & Please think about another situation in which you would like to use the perfect voice assistant.\\ \bottomrule
\end{tabular}%
}
\caption{Scenarios used in our study. Each scenario contains a descriptive part and a specific issue which participants should address and solve in their envisioned dialogue between a user and a perfect voice assistant.}
\label{tab:scenarios}
\end{table*}

\subsection{Scenarios}
We designed eight scenarios based on the most popular use cases for Google Home and Amazon Alexa/Echo as recently identified by Ammari et al. from 250,000 command logs of users interacting with smart speakers~\cite{ammari2019}. In each scenario, we described a specific \textit{everyday situation} a user encounters and an \textit{issue}. Notably, we designed the scenarios in a way so that the participant could choose whether the user \textit{or} the voice assistant initiates the conversation. In addition, we included an open scenario where participants could describe a situation in which they would like to use a voice assistant.
The final scenarios are listed in Table~\ref{tab:scenarios}. This selection of scenarios corresponds to similar analyses of everyday use of voice user interfaces as described by prior research~\cite{cowan2017, bentley2018, luger2016} and consumer reports~\cite{voicereport2020b, voicereport2020}.

\subsection{Procedure}
Participants were introduced to the study purpose and asked for their consent in line with our institution's regulations. 
After that, they were presented with their task of writing dialogues between a user and a voice assistant for different scenarios. We highlighted that the conversation could be initiated by both parties, and also provided an example scenario with two example dialogues (one initiated by the user, the other by the voice assistant). Participants were then presented with the eight different scenarios in random order before concluding with an open scenario, where they were given the opportunity to think of another situation in which they would like to use the perfect voice assistant. For each scenario, participants were asked to first select who is speaking from a dropdown menu (\textit{You} or \textit{Voice assistant}) and then write down what the selected speaker is saying (cf. Figure~\ref{fig:example_dialogue}).
If they wanted, participants could give the voice assistant a name.
At the end of the study, we collected participants' self-reported personality via the Big Five Inventory-2 questionnaire (BFI-2)~\cite{soto2017}, their previous experience with voice assistants, as well as demographic data. 

\subsection{Data Analysis}
\subsubsection{Qualitative Analysis} We conducted a data-driven inductive thematic analysis on the emerging dialogues. Two authors independently coded 27 randomly selected dialogues per scenario (13.2\% of total dialogues), deriving a preliminary coding scheme. Afterwards, four researchers closely reviewed and discussed the resulting categories to derive a codebook. The two initial coders then refined these categories and re-coded the first sample with the codebook at hand to ensure mutual understanding. After comparing the results, the first author performed the final analysis. In case of uncertainty, single dialogues were discussed by two authors to eliminate any discrepancies. In the findings below, we present representative quotes for the themes as well as noteworthy examples of extraordinary dialogues. All user quotes are reproduced with original spelling and emphasis. Our approach follows common practice in comparable qualitative HCI research~\cite{clark2019, cowan2017}. 

\subsubsection{Relationship with Personality}\label{sec:rel_with_perso}
In an exploratory analysis, we analysed the relationship of user personality and the analysed aspects of the dialogues with (generalised) linear mixed-effects models (LMMs), using the R package \textit{lme4}~\cite{Bates2015}. We further used the R package \textit{lmerTest}~\cite{Kuznetsova2017} which provides p-values for mixed models using Satterthwaite’s method. Following similar analyses in related work~\cite{Wu2020}, we used LMMs to account for individual differences via random intercepts (for participant and scenario), in addition to the fixed effects (participants' Big Five personality dimension). In line with best-practice guidelines~\cite{Meteyard2020}, we report LMM results in brief format here, with the full analyses as supplementary material.

\subsection{Participants}
To determine the required sample size, we performed an a priori power analysis for a point biserial correlation model. 
We used G*Power~\cite{faul2009} for the analysis, specifying the common values of 80\% for the statistical power and 5\% for the alpha level. 
Earlier studies regarding the role of personality in language usage~\cite{pennebaker1999, mehl2006} informed the expected effect size of around 0.2 so that we stipulated a minimum sample size of 191. 

We recruited participants using the web platform \textit{Prolific}\footnote{\url{https://www.prolific.co/}, \textit{last accessed 27.07.2020.}}. After excluding three participants due to incomplete answers, our sample consisted of 205 participants (49.3\% male, 50.2\% female, 0.5\% non-binary, mean age 36.2 years, range: 18--80 years). 

Participants on \textit{Prolific} are paid in GBP (\pounds) and studies are required to pay a minimum amount that is equivalent to USD (\$) 6.50 per hour. Based on a pilot run we estimated our study to take 30 minutes. Considering Prolific's recommendation for fair payment, we thus offered \pounds\,3.75 as compensation. We observed a median completion time of 32 minutes with a high standard deviation of 21 minutes. 
Since we wanted to exclude language proficiency and dialect as confounding factors, we decided to only include British English native speakers. 

59.0\% of participants had a university degree, 28.8\% an A-level degree, and 9.8\% a middle school degree (2.4\% did not have an educational degree). 94.1\% of participants had interacted with a voice assistant at least once, while 32.2\% used a voice assistant on a daily basis. Most popular use cases were searching for information and playing music (mentioned by 54.1\% and 51.2\% of participants, respectively), followed by asking for the weather (35.1\%), setting a timer or an alarm (16.1\% and 12.7\%), asking for entertainment in the form of jokes or games (12.7\%), controlling IoT devices (12.7\%), and making a call (11.2\%). Overall, this reflects Ammari et al.'s findings~\cite{ammari2019} on which we based our scenarios. 

\begin{figure}
    \centering
    \includegraphics[width=\minof{\columnwidth}{0.5\textwidth}]{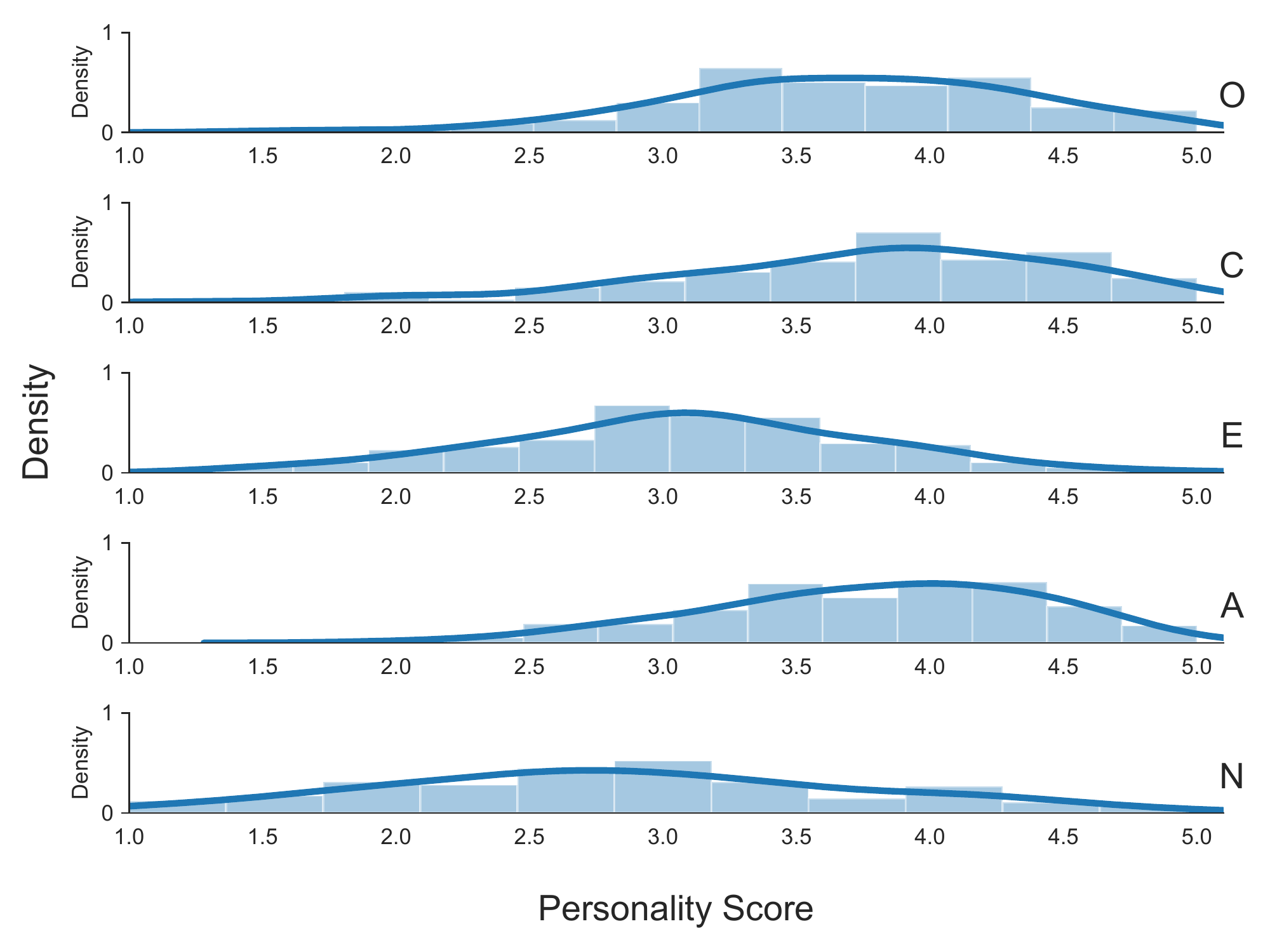}
    \caption{Distribution of the Big Five personality scores in our sample (histogram and KDE plot).}
    \label{fig:big5_distribution}
    \Description[Distributions of the Big Five personality scores in our sample, shown as histograms and KDE plots.]{Distributions of the Big Five personality scores in our sample, shown as histograms and KDE plots. The majority of participants had higher scores in Openness, Conscientiousness, and Agreeableness. Conversely, most participants had values between 2.5 and 3.5 on the Extraversion dimension and lower scores in Neuroticism.}
\end{figure}

Figure~\ref{fig:big5_distribution} shows the distribution of participants' personality scores in the Big Five model.

\section{Results}
We elicited 1,835 dialogues from 205 people with a total number of 81,608 words and 9,282 speaker lines\footnote{Please note that a few participants forgot to indicate the speaker in their dialogues, which we manually added based on dialogue context.}. On average, a dialogue comprised 44.23 words (SD=28.71) and 5.03 lines (SD=2.73).

\begin{table*}[]
    \centering
    \footnotesize
\begin{tabular}{lllllllll}
\toprule
       \textbf{Scenario} & \textbf{Initiated by U} & \textbf{Terminated by U} &      \textbf{Turns U} &        \textbf{Turns VA} &           \textbf{Word count U} &             \textbf{Word count VA} & \textbf{Questions U} & \textbf{Questions VA}\\
\midrule
         Search &            100.0\% &              32.7\% &  3.01 (SD 1.47) &  2.83 (SD 1.45) &  24.53 (SD 12.87) &  31.55 (SD 21.35) & 1.53 (SD 1.01) & 1.47 (SD 1.34) \\
         
          Music &             96.6\% &              28.9\% &  2.37 (SD 1.34) &   2.15 (SD 1.28) &  15.57 (SD 11.99) &  16.02 (SD 14.27) & 0.52 (SD 0.84) & 0.82 (SD 1.06) \\
         
            IoT &             89.3\% &              24.9\% &  2.09 (SD 1.18) &  1.97 (SD 1.19) &   18.91 (SD 12.50) &  16.75 (SD 12.78) &  0.66 (SD 0.78) & 0.56 (SD 0.88) \\
         
         Volume &             72.7\% &              38.0\% &  2.11 (SD 1.23) &   2.06 (SD 1.27) &  16.67 (SD 12.34) &   18.24 (SD 15.01) & 0.66 (SD 0.83) & 0.57 (SD 0.83)\\
        
        Weather &             95.6\% &              41.0\% &  2.73 (SD 1.27) &   2.47 (SD 1.27) &   24.15 (SD 12.59) &   32.40 (SD 21.89) & 1.69 (SD 0.89) & 0.52 (SD 0.74)\\
        
           Joke &             95.6\% &              28.4\% &   2.59 (SD 1.35) &  2.45 (SD 1.36) &  17.69 (SD 11.81) &  23.81 (SD 26.49) & 0.92 (SD 1.02) & 1.10 (SD 1.13) \\
 
 Conversational &             95.1\% &              37.3\% &   2.73 (SD 1.31) &  2.51 (SD 1.31) &  17.88 (SD 10.08) &  23.89 (SD 15.70) & 0.76 (SD 0.93) & 1.05 (SD 0.99) \\
         
          Alarm &             87.3\% &              31.7\% &   2.58 (SD 1.36) &  2.42 (SD 1.32) &   23.70 (SD 14.74) &  23.89 (SD 18.14)  & 0.59 (SD 0.78) & 0.71 (SD 0.85)\\
         
           Open &             95.4\% &              33.2\% &  3.12 (SD 1.63) &  2.98 (SD 1.67) &  25.93 (SD 17.85) &  32.04 (SD 26.18) & 1.04 (SD 1.00) & 1.13 (SD 1.32)\\
\bottomrule
\end{tabular}
    \caption{Automatically extracted data from the dialogues: Percent of dialogues which were initiated and terminated by the user (U) in contrast to the voice assistant (VA). For the other columns, the mean and the standard deviation (SD) over all dialogues in the respective scenario are given.} 
    \label{tab:textanalyticsdata}
\end{table*}

\subsection{Initiating \& Concluding the Dialogue}

\subsubsection{Human vs VA (\textit{Who?})}
We automatically extracted whether a dialogue was initiated by the voice assistant or the user. In the 91.9\% of cases, this was done by the user. Scenario \textit{Volume} presents a notable exception, where the voice assistant initiated 27.3\% of dialogues (cf. Table~\ref{tab:textanalyticsdata}). 
In contrast, overall 67.0\% of the dialogues were terminated by the voice assistant.

\subsubsection{Wake word (\textit{How?})}
When a dialogue was initiated by the user, we analysed whether they addressed the voice assistant by name as a wake word. To this end, we examined whether the first line of a dialogue included the voice assistant name the participant had specified, one of the prevalent voice assistant names (e.g., \textit{Siri, Alexa, Google}), or ``Voice Assistant'' or ``Assistant''. This was true for 73.1\% of user-initiated dialogues.

\subsection{Dialogue Evolution}

\subsubsection{Word count}
As shown in Table~\ref{tab:textanalyticsdata}, the overall number of words (including stop words) varied substantially within one scenario and also differed between scenarios. On average, the voice assistant had a bigger share of speech (M=24.29 words per dialogue, SD=19.09 words) than the user (M=20.56 words per dialogue, SD=12.97 words). 

\subsubsection{Speaker turns}
Despite the smaller share of speech, the user had on average slightly more turns (M=2.59 lines per dialogue, SD=1.35 lines) than the voice assistant (M=2.43 lines per dialogue, SD=1.35 lines). The number of turns did not vary much between scenarios (cf. Table~\ref{tab:textanalyticsdata}). On average, participants described 3.88 speaker turns (SD=2.51 turns) per dialogue. 

\subsubsection{Questioning}
We automatically classified all written sentences as questions vs statements, building on an open source question detection method using the nltk library\footnote{\url{https://github.com/kartikn27/nlp-question-detection}, \textit{last accessed 15.09.2020}}. We further extended this method with a list of keywords that in our context clearly marked a question, as informed by our qualitative analysis (e.g., ``could you'', ``would you'', ``have you''). Table~\ref{tab:textanalyticsdata} shows the numbers of questions per scenario and speaker. Over all scenarios, the grand mean was 0.93 questions for the user and 0.88 for the voice assistant per dialogue.

\begin{figure*}[!t]
\centering
\includegraphics[width=0.75\textwidth]{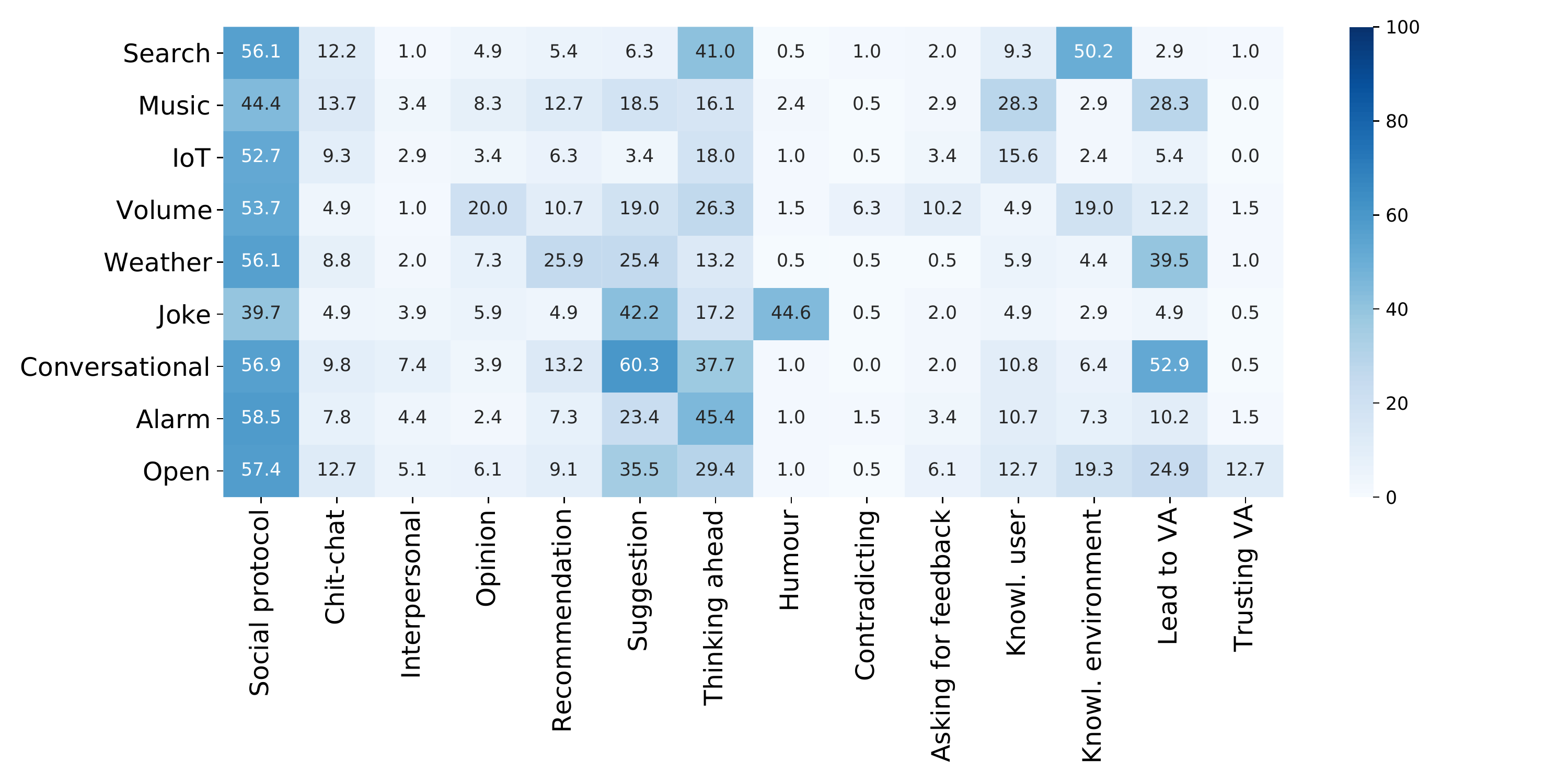}
\caption{Percent of dialogues covering each coded category in each scenario.}
\label{fig:scenario_codings}
\Description[A heatmap matrix, showing the percentages of dialogues covering each coded category in each scenario.]{A heatmap matrix, showing the percentages of dialogues covering each coded category in each scenario. The biggest percentages are for social protocol over all scenarios, suggestion in conversational scenario, thinking ahead in alarm scenario, humour in joke scenario, knowledge about the environment in search scenario, and lead to voice assistant in conversational scenario.}
\end{figure*}

\subsection{Social Aspects of the Dialogue}
Clark et al.~\cite{clark2019} stressed that people perceive a clear dichotomy between social and functional goals of a dialogue with a voice assistant. As anticipated by our study design, the collected dialogues mainly comprised task-related exchange with clear functional goals. Still, the majority of participants also incorporated social aspects, that is, ``talk in which interpersonal goals are foregrounded and task goals -- if existent -- are backgrounded''~\cite{laver1981}. Social talk is not necessary to fulfill a given task, but rather fosters rapport and trust among the speakers and to agree on an interaction style~\cite{dunbar1998}.

Our thematic analysis suggests three different kinds of social talk in the elicited dialogues: \textit{social protocol}, \textit{chit-chat}, and \textit{interpersonal connection}.

\subsubsection{Social protocol} We here define \textit{social protocol} as an exchange of polite conventions or obligations, such as saying \textit{``thank you''}, \textit{``please''}, a form of general affirmation (e.g., \textit{``great''}) or wishing the other a \textit{``good night''}. 91.7\% of participants incorporated at least one of such phrases in at least one of the scenarios. Yet, most participants did not do so in all their dialogues. The use of social protocol ranged from 39.7\% of dialogues in scenario \textit{Joke} to 58.5\% in scenario \textit{Alarm}. 

\subsubsection{Chit-chat} With \textit{chit-chat}, we here refer to an informal conversation on an impersonal level that is not relevant for the actual task.  This includes wishing the user fun or affirming a particular decision (e.g., \textit{VA: ``no problems enjoy the movie i have heard it is very good''} (P40)), assuring to be \textit{``glad to be of service''} (P179), or small talk (e.g., \textit{VA:``Yes, although hopefully will be some sunny breaks in the weather.''} (P105); \textit{``Oh, dinner time, already? Where has the day gone?''} (P55)). 
40.0\% of participants used chit-chat at least once. Chit-chat occurred most frequently in the scenarios \textit{Music} (13.7\% of dialogues), Open (12.7\% of dialogues), and \textit{Search} (12.2\% of dialogues).

\subsubsection{Interpersonal Connection}
Following Doyle et al.~\cite{doyle2019}, we define \textit{interpersonal connection} as talk about personal topics that builds an interpersonal relationship. 20.5\% of participants described interpersonal connection in at least one of the scenarios in a broad range of ways. Interpersonal connection appeared over all scenarios but was slightly more prominent in \textit{Conversational} (7.4\% of dialogues) and \textit{Open} (5.1\% of dialogues). In the former, it was primarily manifested through enquiries about the user (e.g., \textit{VA: ``It appears you are not sleeping yet, what's bothering you?''} (P173)), which the user responds to by sharing what is on their mind, such as anxiety about speaking in public (P145), dealing with a child with autism (P141), or difficulties at work (P87). The voice assistant then comforts the user (e.g., \textit{``Don't worry I got the perfect plan''} (P120)) or makes suggestions on how to deal with the situation. For example, P87 sketched a voice assistant which offers to have the user's back (e.g., \textit{U: ``My boss reprimanded me'' -- VA: ``WHAT?? Shall I suggest ways to take your revenge? [...] Take me into work with you with your headpiece on and I'll suggest replies the next time he's nasty to you''}), and P152's voice assistant motivates the user in a witty way (e.g., \textit{VA: ``Get out of bed and then i will start'' -- U: ``That's harsh'' -- VA: Come on, i'll play the Spice Girls if you promise to dance along and sing into your hairbrush''}). 

In the \textit{Open} scenario, interpersonal connection was manifested in various ways, such as through emphasising the relationship with the user (e.g., \textit{VA: ``I hope you wouldn't ever lie to me as I'm your best friend''} (P87)), by recollecting shared experiences (e.g., \textit{VA: ``Here are my favorite pictures of last Halloween. Personally this was my favorite costume, and if I remember correctly we listened to this artist all night. I turn up the music and play some of her tunes.''} (P2)), or discussing the user's love life (P179): 
\begin{quote}
VA: \textit{``Was that hesitation I registered in your voice?''}

U: \textit{``No, what are you talking about? Of course I'm cooking for myself, who else would I be cooking for?''}

VA: \textit{``A lady, maybe? ''}

U: \textit{``.....''}

VA: \textit{``What's her name?''}

U: \textit{``None of your business''}

VA: \textit{``Dude I am an AI that lives in your house, of course it's gonna be my business. If it is a lady coming over then you need to be a lot cooler than you are with me''}

U: \textit{``Ahh dude you're right, I'm sorry I'm just nervous''}

VA: \textit{``No shit''}
\end{quote}
 
\subsection{Voice Assistant Behaviour}
Our thematic analysis showed that the majority of participants let their voice assistant take the lead in parts of the dialogue. By \textit{taking the lead}, we refer to the voice assistant either providing advice to the user or doing something the user did not specifically ask for. We further differentiate between \textit{suggesting}, \textit{recommending}, \textit{giving an opinion}, \textit{thinking ahead}, \textit{contradicting}, \textit{refusing}, \textit{asking for feedback}, and \textit{humour}, as emerged from our analysis.

\subsubsection{Suggesting \nopunct} denotes ``mention[ing] an idea, possible plan, or action for other people to consider'' according to the Cambridge Dictionary\footnote{https://dictionary.cambridge.org/dictionary/english/suggest, \textit{last accessed 12.09.2020.}}. That is, the voice assistant selects individual options and presents them to the user, without indicating a preference for one of them. Suggestions are often introduced by 
\textit{``What about''}, \textit{``You could''}, or \textit{``Which would you prefer''}. 87.8\% of participants had their voice assistant give at least one suggestion over all dialogues. Suggestions occurred most often in the scenarios \textit{Conversational} (60.3\% of dialogues) and \textit{Joke} (42.2\% of dialogues), while less than 10\% of dialogues in the scenarios \textit{Search} and \textit{IoT} contained a suggestion.

Suggestions mainly came in the form of possible options the voice assistant pointed out to the user, such as films, music, books, jokes, games, or recipes. When giving a suggestion, the voice assistant often took into account user preferences (e.g., \textit{VA: ``There's a film called Onward from Disney, I know you like the Pixar films.''} (P107)) or context (e.g., \textit{VA: ``What are you cooking today?'' -- U: ``I'm making meatloaf.'' -- VA: 
``OK, I've found a playslist for you starting with Bat out of Hell.''} (P67)).

Other suggestions were more complex. Depending on the scenario, this included strategies for falling asleep, avoiding oversleeping, preparing a trip, or dealing with the neighbours (\textit{e.g., U: ``Hey Lexi, My neighbours think my music is too loud.'' -- VA: ``How about I find a new home?''
-- U: ``No, that isn't realistic enough. -- VA: ``What if i search for some great headphones?'' -- U: ``Sure! That would be great'' -- VA: ``I will get onto that.''} (P27)).

\subsubsection{Recommending,\nopunct} on the other hand, describes advising someone to do something and emphasising the best option\footnote{https://dictionary.cambridge.org/dictionary/english/recommend, \textit{last accessed 27.07.2020.}}. The voice assistant usually ushers recommendations by phrases such as \textit{``I would do/choose''} or \textit{``I recommend''}. 51.7\% of participants let their voice assistant give at least one recommendation in varying complexity. For instance, the advice given by P8's voice assistant is rather straightforward (\textit{``I recommend spaghetti ala carbonara''}) while P15 described a more complex recommendation: \textit{VA: 
``hey rami, let me optimise the frequency of the speakers so we have the maximum volume indoors without decibels spilling over into the neighbours ear shot.'' -- U: ``that's great, I didn't even know you could do that''}. Recommendations concerned entertainment, such as films, music, videos, or how users could achieve their goals, for example, going to the cinema or not to oversleep. In a few cases, the voice assistant nudged the user towards better behaviour (e.g., \textit{VA: ``You should turn it [the music] down as your neighbours have complained before''} (P186)) or helped saving energy (P103):
\begin{quote}
    VA: \textit{``I've noticed you've been leaving the lights on all night.''}
    
    U: \textit{``I know. It's when I read in bed. I fall asleep and forget to turn them out.''}
    
    VA: \textit{``Did you want me to turn them off for you? [...]''}

    U: \textit{``Will it make much difference whether they are on or off?''}
    
    VA: \textit{``It'll save electricity. On your current plan, you would save £3.00 per month by turning out the lights every night.''}
\end{quote}

Over all scenarios, recommendations occurred most prominently in \textit{Weather} (25.9\% of dialogues), in which the voice assistant recommended what to pack. These recommendations were often based on knowledge about the weather forecast (e.g., \textit{VA: ``The weather in Rome, Italy is expected to be hot and dry this week. I would recommend bringing light, breathable shorts and shirts.''} (P137)) or knowledge about the expected context (e.g., \textit{VA: ``The best views of the city are from the gardens above the valley, so make sure to take something you can walk comfortably in.''} (P125)).  

Second was the \textit{Conversational} scenario (13.2\% of dialogues), in which the user asked the voice assistant's help for falling asleep (e.g., \textit{U: ``I cannot sleep, do you have any useful recommendations?''} (P20)).

\subsubsection{Giving an Opinion\nopunct} refers to sharing thoughts, beliefs, and judgments about someone or something\footnote{https://dictionary.cambridge.org/dictionary/english/opinion, \textit{last accessed 27.07.2020.}}. 39.0\% of participants had their voice assistant give an opinion in at least one dialogue. In terms of scenarios, the voice assistant most often expressed an opinion in \textit{Volume} (20.0\% of dialogues). Here, the voice assistant commented: 
\textit{``I think the music you are playing is too loud. It will annoy your neighbours.''} (P204). Apart from this, the voice assistant also shared its opinion on the user's choice of film or food, usually praising the user (e.g., \textit{``U: Great, I think I'd like to go to see (film) at 7pm.'' -- VA: ``Good choice [...]''} (P191); \textit{U: ``Hey Lexi, I'm going to Italy'' -- VA: ``Ciao! what a beautiful country''} (P27)). Moreover, it commented on bad habits of the user, such as \textit{VA: ``Haha yeah it's [leaving on the lights at night] not a good habit''} (P179). In addition, the voice assistant shared its taste, confident that the user will like it, too: \textit{VA: ``I'll play you a mix of some songs you know and some new things I think you'll like.''} (P48).

\subsubsection{Thinking ahead \nopunct} describes that the voice assistant anticipates and proposes possible next steps to the user without the user asking for them. Note that this does not include voice assistant enquiries due to incomplete information on a task (e.g., \textit{U: ``Hi frank, what are the film times for local cinema'' --  VA: ``Please choose which cinema''} (P45)). Examples for \textit{thinking ahead} include offering to book tickets when a user asks for film showings (e.g., \textit{VA: ``They play it [the film] at 7:30pm on Saturday. Do you want me to book it?''} (P13)), suggesting to set a reminder or a morning routine (e.g., \textit{VA: ``No problem I will wake you at least one hour before that and prepare a coffee so that you are actually awake.''} (P2)), or making the user comfortable (e.g., \textit{VA: ``Its going to be chilly in the morning shall I set the Hive for the heating to come on a little earlier than usual so its warm when you get up?'' (P99))}. 83.4\% of participants created such a foresighted voice assistant at least once, even though their users did not always accept the proposed actions. Thinking ahead was particularly prevalent in the scenarios \textit{Alarm} (45.4\% of dialogues) and \textit{Search} (41.0\% of dialogues). 

\subsubsection{Contradicting \nopunct} denotes parts of the dialogue in which the voice assistant disagrees or argues with the user. Only 8.3\% of participants let the voice assistant contradict the user at least once. Single cases of contradicting were spread over all scenarios, while most occurrences were part of the scenario \textit{Volume} (in 13 out of 205 dialogues). While in some cases, the voice assistant carefully phrased its objection (e.g., \textit{U: ``I don't think so Sally, I like it loud.'' -- VA: ``Well, forgive me, but I have very sensitive hearing and can hear them next door getting a bit upset with you.''} (P65)), it made this objection very clear in others (e.g., \textit{VA: ``You do realise that your music is so extremely and inconsideratly loud that it could be annoying everyone including the neighbours''} (P106)). Other examples for contradicting included the voice assistant having a different opinion on a particular topic (e.g., \textit{U: ``Hmm, no I don't like her [Jennifer Anniston] as an actress.'' -- VA: ``She is very talented''} (P20) or on fulfilling a task (e.g., \textit{U: ``Can you set the alarm for 8am please?'' -- VA: ``Maybe I should set it for 7.30am just in case and to give you more time.''} (P20)). Interestingly, in all arguments, the user always gave in and followed the voice assistant's advice.

\subsubsection{Refusing}
Only three participants (in four dialogues) had the voice assistant refuse what the user asked for. For example, the voice assistant declined to increase the volume of the music (\textit{VA: ``Yes, but it's so loud that it's keeping me awake''} (P87)) and asked the user instead to \textit{``please plug in your headphones''}. Another participant (P179) described an assertive human-like voice assistant which tells the user's friends a funny story about the user despite their protest, and fights with the user about who turns down the music: \textit{VA: ``You're the one with hands, you turn it down'' -- U: ``You're literally in the ether where the electronics live, you turn it down [...]''}.

\subsubsection{Asking for Feedback} 
21.5\% of participants let the voice assistant ask the user for feedback on how well it had done or for confirmation to proceed. Asking for feedback was not particularly prominent in any of the scenarios but was described most often in \textit{Volume} (10.2\%), for instance, to enquire if the user was \textit{``happy with that [adjusted] sound level''} (P21). In other scenarios, the voice assistant wondered whether the user \textit{``[liked] that story''} (P81) or the music (\textit{``Ok, but if it is getting too funky just say it!''} (P2)) the voice assistant had suggested. 

\subsubsection{Humour}
45.4\% of participants adorned their voice assistant with humorous statements and context-aware, pragmatic, and funny comments. Unsurprisingly, most occurred in the \textit{Joke} scenario (44.6\% of dialogues). 

Examples from other scenarios included remarks on the user's film choice \textit{Terminator} (\textit{VA: ``and don't forget, I'll be back''} (P19)) 
and sarcasm when asked for a suitable conversation topic (\textit{VA: ``Ok. Lets make it interesting. What's everyone's position on brexit?''} (P113)). However, voice assistant humour appealed to the users differently. While some dialogues encompassed appreciation (e.g., \textit{U: ``You are so funny''} (P28)), others described the user as less convinced (e.g., \textit{U: ``Nice try''} (P244)).

\subsection{Voice Assistant Knowledge}
Participants attributed the voice assistant \textit{knowledge about the user} as well as \textit{knowledge about the environment}.

\subsubsection{Knowledge about the User} With \textit{knowledge about the user}, we refer to voice assistant knowledge about user behaviour and preferences. For example, when the voice assistant is aware of the user's schedule (e.g., \textit{VA: ``It looks like the 8PM showing would fit into your schedule best''} (P178)) and favoured choices (e.g., \textit{VA: ``Maybe one of your favourite playlists - last time you were cooking you played this one?''} (P99)). Participants also let the voice assistant know about the user's health (e.g., \textit{VA: ``I see your heart beep is moving irregular[ly]. You okay[?]''} (P120)), habits (e.g., \textit{U: ``Hey Masno, you know i snoring every night.'' -- VA: ``Yes, you are so loud.''} (P5)), and past events (e.g., \textit{VA: ``Hi, It's Sally, why are you not sharing the stories about your last holiday with your friends?''} (P65)). 58.5\% of participants equipped the voice assistant with knowledge about the user in at least one scenario. In terms of the scenarios, this kind of knowledge was most strongly represented in \textit{Music} (28.3\% of dialogues) and \textit{IoT} (15.6\% of dialogues), where it is primarily related to the user's preferences in terms of musical taste. Conversely, in scenario \textit{IoT} the voice assistant is equipped with knowledge about the user behaviour, in particular to automatically recognise whether they are already asleep.

\subsubsection{Knowledge about the Environment}
Knowledge about the environment includes intelligence about the status of other devices in the house (e.g., \textit{U: ``Henry, can you tell me what's low in stock in the fridge[?]''} (P185)) as well as the ability to interact with these devices (e.g., \textit{VA: ``I will get the coffee machine ready for when you wake so the smell might get you to rise'' (P105)}). It also comprises awareness of the current location and distance to points of interest in the vicinity (e.g., \textit{U: ``Hey Masno, could you check whats time the local cinema are playing this new action film?'' (P5})), and a kind of omniscient knowledge about others (e.g., \textit{VA: ``I'll turn it down when I hear them enter th[ei]r house.'' (P23))}. 71.2\% of participants equipped the voice assistant with such knowledge in at least one of the scenarios. This was most prevalent in \textit{Search} (50.2\% of dialogues) with knowledge about the nearest or local cinema, followed by \textit{Volume} (19.0\% of dialogues) with knowledge about the neighbours, and by the \textit{Open} scenario (19.3\%). In the latter, the voice assistant could often tell its user what is in the fridge, interact with other devices in the house, or even knew the stock and prices of items in all local supermarkets (e.g., \textit{U: ``Can you check my local supermarkets to see if anyone has got Nescafe on offer?'' -- VA: ``I can see that Morrisons has 2 jars for the price of 1, would you like me to add this to your shopping list?''}).

\subsection{User Behaviour}
\subsubsection{Trusting the Voice Assistant with Complex Tasks}
16.1\% of participants trusted the voice assistant to execute demanding social or complex tasks appropriately without detailed instructions, which, if executed incorrectly, could have negative social or professional repercussions. The \textit{Open} scenario recorded the most occurrences of this category (12.7\% of dialogues), indicating that trusting the voice assistant with challenging tasks is more of a future use case. Examples included social tasks, such as writing a message without specifying the exact content (e.g., \textit{U: 
``I need you to write an email to my daughter's college. [...] The additional help provided for her because of her dsylexia. They promised reader pens and a dictaphone but she hasn't received them yet. Please ask why''} (P41)), sending out birthday cards, or selecting pictures to show to friends. Moreover, participants trusted the voice assistant with preparing a weekly meal plan and ordering the according ingredients, putting together a suitable outfit or planning a trip, paying for expenses, or editing presentations for work as well as making a website.

\subsubsection{Giving the Voice Assistant the Lead} 80.0\% of participants let the user hand over the lead of the dialogue to the voice assistant at least once, for example by asking for an opinion or recommendation. The occurrence of this category varied greatly across the scenarios and was most pronounced in \textit{Conversational} (52.9\% of dialogues) and \textit{Weather} (39.5\% of dialogues). In the former, participants sought advice from the voice assistant on how to fall asleep or waited for the voice assistant to help by simply stating that they were 
\textit{``having trouble falling asleep'' (P11)}. In \textit{Weather}, participants let the user not only ask the voice assistant for the weather but also for recommendations on what to pack (e.g., \textit{``U: Can I ask your advice on what type of clothes to pack for Italy, will I need any light jumpers or anything?''} (P107)). Participants also liked to see the voice assistant as a source of inspiration, which provides suggestions on what to read, cook, play, or listen to. For example, P2 requested: \textit{``Surprise me and play something you like''}.

\subsubsection{Assigning Characteristics to the VA} 
In a few cases, participants incorporated an explicit description of the voice assistant by letting the user comment on them (e.g., \textit{``U: You are so funny.''} (P103)). These descriptions were only found in sixteen dialogues (0.9\%) and included seven times \textit{``funny''}, four times \textit{``smart''} or \textit{``clever''}, and once \textit{``reassuring''}. Once the voice assistant was called an \textit{``entertainer''} and a \textit{``mindreader''}. One participant noted: \textit{``You’ve got my back -- ain’t you''} (P179). %
On the other hand, three participants also commented on the voice assistant's lack of wittiness (e.g., \textit{``No! Something actually funny!''} (P103)).

\subsection{The Status Quo}
Finally, we analysed how many dialogues did \textit{not} fall into any of the aforementioned categories. These dialogues can be seen as a depiction of the status quo: a functional task-related request. Table~\ref{tab:status-quo} provides example status quo dialogues for each scenario. The occurrence of these dialogues ranged from 14.2\% in scenario \textit{Joke} to 32.7\% in scenario \textit{IoT}.

\begin{table*}[]
    \centering
    \footnotesize
\begin{tabular}{lll}
\toprule
       \textbf{Scenario} & \textbf{\% of dialogues} & \textbf{Example Status Quo Dialogue}\\
\midrule
Search & 20.0\% & U: \textit{``Assistant, search up the film times for Shrek at the Odeon in Liverpool.''} -- 
VA: \textit{``The film times are at 2:00, 2:45 and 5:00.''} (P170) \\

Music & 22.4\% & U: \textit{``Eleonora play Tom Petty on Spotify.''} --
VA: \textit{``Playing songs by Tom Petty on Spotify.''} (P26) \\

IoT & 32.7\% & U: \textit{``Google, switch off all home lights at 2am.''} -- VA: \textit{``Ok, done, lights will switch off at 2am''} (P34) \\

Volume & 27.8\% & U: \textit{``Alexa turn the music down to 6.''} --
VA: \textit{``Ok.''} (P18) \\

Weather & 21.0\% & U: \textit{``Minerva, what is the weather going to be like in Italy this week?.''} --
VA: \textit{``The weather will be mostly sunny in Italy this week.''} (P186)\\

Joke & 14.2\% & U: \textit{``Bubble, it's a party! Tell us something fun and interesting.''} --
VA: \textit{``Here are some fun stories i have found on the internet..''} (P109) \\

Conversational & 16.7\% & U: \textit{``Hey google, play rain sounds.''} -- VA: \textit{``Playing rain sounds''} (P6) \\

Alarm & 24.4\% & U: \textit{``Google, set an alarm for 8am.''} --
VA: \textit{``OK , alarm set for 8am.''} (P34) \\

Open Scenario & 21.3\% & U: \textit{``Dotty, reminder for hospital appointment at 3 pm tomorrow.''} --
VA: \textit{``Reminder set.''} (P15) \\ 

\bottomrule
\end{tabular}%
\caption{Example dialogues for each scenario which did not fall into any other category. These dialogues can be seen as a depiction of the status quo: functional task-related request. Percentages refer to their share in all dialogues per scenario.}
\label{tab:status-quo}
\end{table*}

Status quo dialogues were on average shorter than dialogues overall (difference between the two indicated by $\Delta$, respectively). User (M=12.15 words per dialogue, SD=8.05, $\Delta$=8.41) and in particular voice assistant (M=10.01 words per dialogue, SD=9.82, $\Delta$=14.28) had a smaller share of speech in the status quo dialogues in contrast to the average word count over all dialogues. 
Similarly, there were fewer speaker turns both by the user (M=1.57, SD=0.85, $\Delta$=1.02) and the voice assistant (M=1.56, SD=0.88, $\Delta$=0.87). 
97.2\% of the status quo dialogues were initiated by the user, while 89.2\% of the status quo dialogues were terminated by the voice assistant.

\subsection{Open Scenario}
As a last task, participants were asked to write a dialogue for another scenario they would like to use their perfect voice assistant in. These dialogues indicated a broad spectrum of imagined use cases, yet the majority (53.8\%) reflected already existing ones. This is common when people are asked to imagine a technology which does not exist yet~\cite{tabassum2019}. 
These scenarios included receiving recommendations or suggestions from the assistant (mentioned by 11.2\% of participants in all open scenarios), searching for information (11.2\%), controlling IoT devices (10.7\%), using the assistant instead of typing (e.g., for notes, shopping lists, text messages; 9.6\%), getting directions (6.1\%), or setting an alarm, timer, or notification (3.0\%).

However, 44.7\% of people mentioned scenarios in which the voice assistant's capabilities exceed the status quo. In most of these cases (45.7\%), they imagined the voice assistant to become a personal assistant with very diverse roles and tasks, which supports them in their decision-making. For example, P114 would like to have cooking assistance: \textit{``I would ask the voice assistant [...] for help in cooking dishes like homemade curries and perfect pork crackling joints and perfect roast potatoes''}. P65 saw her perfect voice assistant as a diet and meal planner: \textit{``[It orders] me food shopping with good dates, healthy choices in the foods I like. It would also consider my dietary requirements (lactose intolerant) and add the substitutes I like for dairy [...]''}. P13 imagined a housework organiser (\textit{``To plan my housework for the week and give me reminders to do it. And chase me up if I don't say that it is completed''}), and P61 a personal shopper (\textit{``Give your preferences, size etc [...] Give [...] the event type you are attending and your price range and ask to order you outfits for the occasion.''}). 
P139 even trusted the voice assistant in \textit{``coping with an autistic child and helping to handle them''}, and P25 would like to use it for mental health support. Another four people described a scenario in which the assistant helps in an emergency, such as alerting the neighbour in case of a household accident (P73).

We further classified the roles participants implicitly ascribed to their perfect voice assistant in the \textit{Open} scenario. Three different roles emerged from the dialogues: \textit{tool}, \textit{assistant}, and \textit{friend}. We classified the role as \textit{tool} if the user utilises the voice assistant in order to do something they want to do\footnote{https://dictionary.cambridge.org/dictionary/english/tool, last accessed 04.01.2021}, that is, a clearly defined task which the voice assistant simply carries out. 26.9\% of all \textit{Open} scenario dialogues featured a voice assistant as a tool. For example, P38 sketched a dialogue for setting an alarm: \textit{U: ``Set an alarm for 10 minutes please.'' - VA: ``Alarm set''}. A voice assistant as an \textit{assistant} is someone who helps the user to do their job\footnote{https://dictionary.cambridge.org/dictionary/english/assistant, last accessed 04.01.2021}. In contrast to the tool, however, the task is not precisely defined, but requires a certain amount of creativity, thinking ahead, or individual responsibility. Moreover, the voice assistant is seen as a person rather than a thing. 71.6\% of participants ascribed an assistant role to the voice assistant. For example, P16 would like support to find presents: \textit{U: ``Hey google, it's sarah's from works birthday on 22nd January, can you remind me to get her a gift?'' - VA: ``Hey rami, sure thing, let me put that in the calendar for you. We can put together a list of gift ideas, do you have anything in mind?''}
Finally, a voice assistant as a \textit{friend} knows the user well and has a close, personal relationship with them\footnote{https://dictionary.cambridge.org/dictionary/english/friend, last accessed 04.01.2021}. Only three participants imagined a closer relationship with their voice assistant -- \textit{``a best friend who will never betray me. :-)''}, as P86 put it.

\subsection{Relationship with Personality}

\begin{figure*}[!t]
\centering
\includegraphics[width=0.8\textwidth]{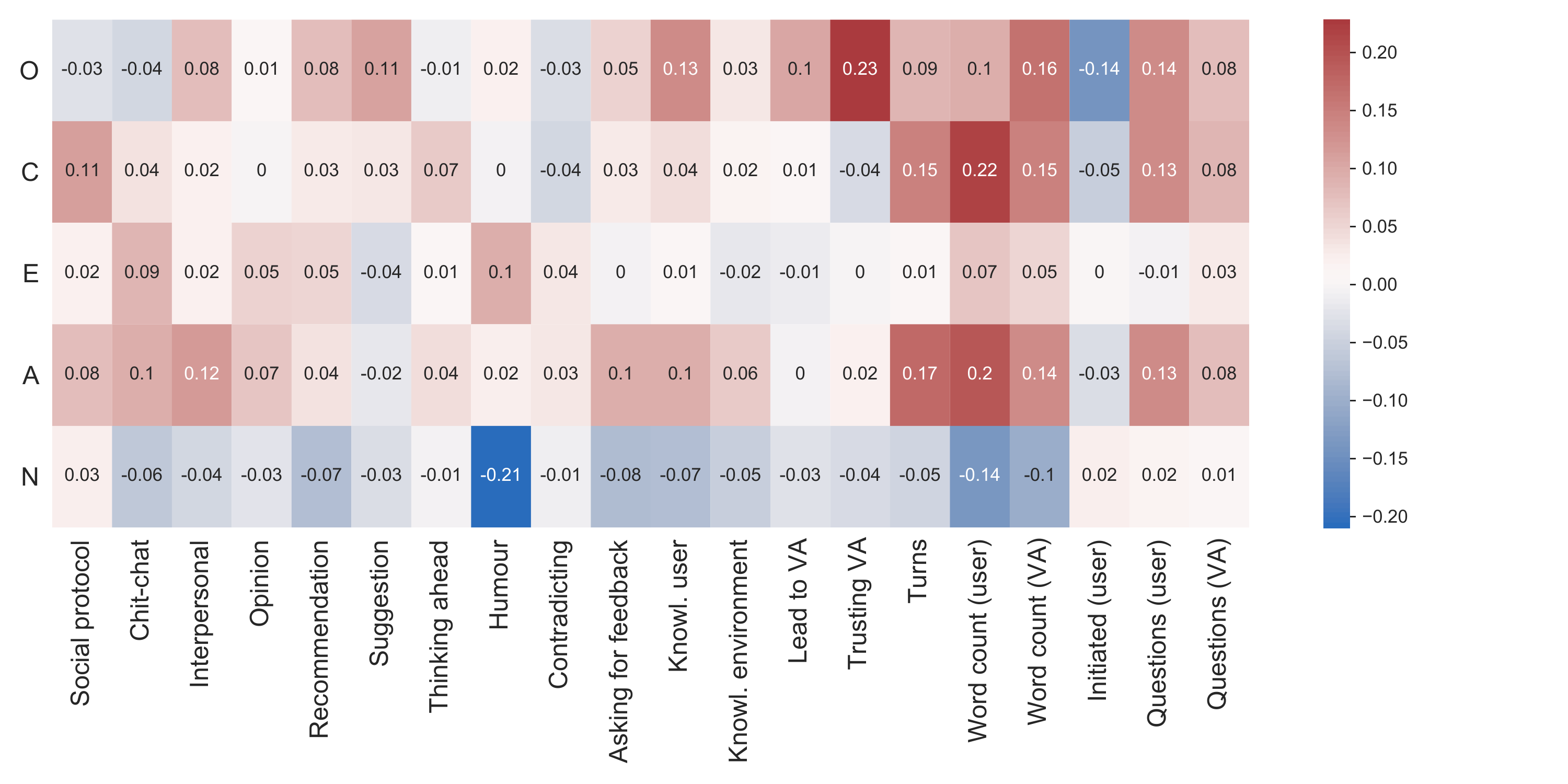}
\caption{Spearman correlations of Big 5 personality scores and aspects of the dialogues.}
\Description[A heatmap matrix, showing the percent of dialogues covering each coded category in each scenario.]{A heatmap matrix, showing the percent of dialogues covering each coded category in each scenario. Most prominent were Suggestions in the Conversational scenario with 60\%, Thinking ahead in the Alarm scenario with 45\%, Humour in the Joke scenario with 44\%, Knowledge about the environment in the Search scenario with 50\%, and Lead to the VA in the Conversational scenario with 53\%.}
\label{fig:correlations}
\end{figure*}

As an overview, Figure~\ref{fig:correlations} shows the correlation coefficients between user personality and the examined measures. We overall see positive associations of \textit{Conscientiousness}, \textit{Openness}, and \textit{Agreeableness} with measures of dialogue length (turns, word counts of both user and voice assistant). Further associations stand out for \textit{Openness} and \textit{Trusting the VA} (positive), and \textit{Neuroticism} and \textit{Humour} (negative).

In addition, we created one generalised LMM for each measure, as described in Section~\ref{sec:rel_with_perso}. Since the \textit{Open} scenario dialogues highly depended on the individual use case, we excluded this scenario from the analysis. 
For brevity, we only report on some of the models here. In particular, to account for the exploratory nature of our analysis, we make this decision based on the \textit{uncorrected} p-value: That is, we report on all models with a predictor with p<.05. We provide the analysis output of all models in the supplementary material. Since this is an exploratory analysis, we highlight that significance here is not to be interpreted as confirmatory. Rather, we intend our results here to serve the community as pointers for further investigation in future (confirmatory) work.

For \textit{Opinion}, the model had \textit{Conscientiousness} as a significant negative predictor (\glmm{-0.483}{0.239}{-0.359}{-0.951}{-0.014}{-2.01}{<.05}), indicating that people who score higher on this personality dimension might prefer voice assistants that less frequently express own opinions. Based on the coefficient $\exp(\beta_{std})$, a one point increase in \textit{Conscientiousness} results in 0.70 times the chance of including an opinion in the dialogue.

For \textit{Humour}, the model had \textit{Neuroticism} as a significant negative predictor (\glmm{-0.742}{0.250}{-0.664}{-1.232}{-0.253}{-2.97}{<.01}), indicating that people who score higher on this dimension might prefer assistants that less frequently express humour: In this model, a one point increase in \textit{Neuroticism} results in 0.51 times the chance of including humor in the dialogue.

For \textit{Question (User)}, the model had \textit{Conscientiousness} as a significant positive predictor (\glmm{0.176}{0.088}{0.133}{0.005}{0.348}{2.01}{<.05}), indicating that people who score higher on this dimension might prefer asking more questions when conversing with voice assistants: In this model, a one point increase in \textit{Conscientiousness} results in 1.19 times the chance of a user's sentence being a question.

\section{Limitations}
Our data, method, and findings are limited in several ways and should be understood with these limitations in mind. 

First, while our scenario selection was informed by the most popular real-world use cases for voice assistants~\cite{ammari2019}, our data was collected in an online survey. In contrast to everyday use of voice assistants, where conversations are usually embedded in various real-life situations, this created a more artificial setting which might have influenced the dialogue production~\cite{porcheron2018}.
Moreover, users might display different dialogue preferences in practice than in theory.  

Second, our dialogues were written down, and are therefore limited in what they can tell us about actual, spoken conversations. This concerns, for example, the negotiation of turn-taking, which is usually an important part of conversation analysis~\cite{sacks1974}, but cannot be assessed on our data. We focus on how the envisioned dialogues should be structured in terms of content and proportion of distinct linguistic behaviour (e.g., contains social talk). Paralinguistic aspects of speech (e.g., accent, tone) are equally important, but require spoken conversation. However, it might be more difficult for participants to embody both voice assistant and user while inventing a spoken dialogue in a study. Evaluating the content -- as we did -- may therefore be (initially) more actionable for user-centred design. Also, asking crowdworkers to write dialogues for a conversation flow has been effectively utilised before~\cite{choi2020}, further demonstrating the potential for written dialogue elicitation in conversational interface design.

Third, as anticipated by our study design, the collected dialogues mainly comprised task-related conversation with clear functional goals. Our findings and implications thus might not generalise to non-task-related dialogues with a fuzzy goal or no goal at all.

Fourth, while writing the dialogues, participants had to anticipate a technology which does not yet exist, at least in the form we requested. We acknowledge that the dialogues might therefore have been influenced by participants' imagination. More creative participants might have come up with richer dialogues than others. 

Finally, it is important to note that, rather than representing gold standard voice assistant interactions, the dialogues here should be interpreted as the first step in a user-centred design process towards personalised dialogues based on users’ vision of what perfect voice assistants should do in these tasks. As such, the elicited dialogues can inform the design of personalised voice assistant prototypes in a next step. However, as in any user-centred design process, these dialogues and prototypes must then be evaluated and validated with users. In particular, users' visions of a perfect voice assistant might change after experiencing the use of such a voice assistant. It is therefore essential to understand the design of personalised voice assistants as a process with several iteration loops.

\section{Discussion}
By writing their envisioned dialogues with a voice assistant, participants implicitly painted a picture of the characteristics of their perfect voice assistant. In the following subsection, we analyse and discuss these characteristics. 

\subsection{What or Who is the Perfect Voice Assistant?}
Our first research question asks how users envision a conversation with a perfect voice assistant. The wide range and diversity of dialogues suggest that there is no single answer. Here, we discuss both common trends as well as diverging preferences. 
Moreover, we point out implications for voice assistant design and research.

\subsubsection{Smarter, More Proactive, and Equipped with Personalised Knowledge} The majority of people envisions a voice assistant which is smarter and more proactive than today's agents, and which has personal knowledge about users and their environment. In particular, it gives \textit{well thought-through suggestions and recommendations} to solve complex problems.

The perfect assistant is also \textit{foresighted and proactive}, anticipating possible next actions. However, users in the dialogues do not always accept their assistant's suggestions. Together, these findings indicate that, rather than a master-servant relationship~\cite{doyle2019}, users wish for perfect voice assistants to be more collaborative.

\textit{Knowledge about the users and their environment} may also make conversations with assistants more effective and natural by creating the impression of shared knowledge and common ground, as integral to human dialogue effectiveness~\cite{clark2019, clark1996using}. 

Such shared knowledge is currently missing in the design of voice assistants~\cite{clark2019}. Considering our results on questions, interactivity, and ``thinking ahead'', this might be realised in current systems by allowing the assistant to proactively ask the user (more) questions at opportune moments. Moreover, knowledge about the user also allows for more personalised suggestions and conversations, which are more likely to appeal to the user. 

\subsubsection{More than Fast Information Retrieval}
Another trend in the majority of dialogues is that they are not intended or optimised for fast information retrieval. Current dialogues with voice assistants are characterised by a question-answer structure~\cite{doyle2019} and a median command length of four words~\cite{bentley2018}. In contrast, people's envisioned dialogues comprise longer speech acts and more interactivity, creating the impression of being more \textit{conversational}. This is further supported by the observed amount of non-task related talk, such as chit-chat, personal talk, or humour. Hence, it appears as if there is a demand for more human-like personal conversation with voice assistants than currently available despite recent discussions whether humanness is the best metaphor to interact with conversational agents~\cite{doyle2019}. 

In the long term, the design of voice assistants should aim for multiple-turn conversations. Yet, in the short term, a variety of fillers to begin answers (e.g., \textit{``Sure, let me get on to this.''}) and closing remarks (e.g., \textit{``Enjoy the movie!''}) could be used to avoid raising unrealistically high expectations.

\subsubsection{A Range of Roles: Tool, Assistant, or Friend?}
People imagine different roles for their perfect voice assistant: 22\% of dialogues were purely functional, suggesting that the assistant is seen as a tool to get things done. However, the majority of dialogues depicts a helpful assistant who supports the users in their chores and might take over more complex tasks in the future, as suggested by participants in the \textit{Open} scenario. As a consequence, users feel obliged to obey to conversational rules, including ``thank you'' and ``please''. The number of participants following these social protocols was higher than expected from previous studies examining interaction with a robot receptionist~\cite{lee2009}. It was also surprising that 40.0\% of participants included a form of chit-chat since this kind of small talk was previously flagged as inappropriate and unwelcome~\cite{doyle2019, wilks2010}. A reason for this difference to related work could be that participants imagined a more intelligent voice assistant than is currently available. 
Echoing previous findings~\cite{doyle2019, clark2019},  few participants regarded the voice assistant as a \textit{friend}. However, the scenarios participants described in the \textit{Open} task reveal use cases in which the assistant also listens to and advises on personal issues.

Overall, these findings motivate considering such \textit{roles} as a conceptual basis for personalisation of voice assistants, beyond or in addition to the currently dominant focus on personality. For instance, people living alone or with a smaller circle of acquaintances might be more likely to seek personal advice from a voice assistant, seeing it as a friend rather than an assistant.

\subsubsection{Emancipated vs Patronised}
39.0\% of participants designed an emancipated voice assistant which expresses its own opinions, and 8.3\% even allowed it to nudge them to behave in a certain way. 
This contradicts prior work by~\citet{doyle2019}, who found that speech agent users were suspicious of the agent expressing an opinion. A reason for this discrepancy could be that today's voice assistants often do not live up to user expectations~\cite{luger2016, cowan2017, porcheron2018}. However, being interested in and accepting someone's opinion requires a certain level of trust in their skills, knowledge, and experience. Hence, our findings point to a mixed picture, in which some users appreciate the voice assistant's advice provided they perceive it has the required skills to give a useful opinion. As P143 puts it, \textit{``[w]e all need a bit of help from time to time and advice, and yet sometimes there is nobody to talk to. The option to ask [for] an opinion would be a great thing to have at anybody's disposal.''}
In addition, most participants did not link their voice assistant to a particular company, which might also influence trust in its opinions. In certain situations, a small part of participants seems to accept a voice assistant which contradicts the user. In one of our scenarios, participants heeded the voice assistant's objection to avoid a conflict with the neighbours. Thus, future work could leverage this knowledge and evaluate the effectiveness of persuasive voice assistants for other topics such as in supporting a healthy lifestyle or environmental-friendly behaviour. In the short term, a voice assistant could offer its ``own'' opinions from time to time, yet only after the user has asked it to do so at least once.

\subsubsection{The Thing about Humour}
Humour is considered an integral part of conversations with humans as well as an interesting novelty feature and entry point for voice assistants~\cite{clark2019, luger2016}. Our findings suggest that there are individual preferences for humour: More than half of the participants did not equip their perfect assistant with a sense of humour although they were given the task to entertain their friends. Three people even let the user comment on the voice assistant's lack of ``actual'' humour. On the other hand, others acknowledged the assistant's wittiness. Apart from the \textit{Joke} scenario, humour was often included in the form of comments on the situation, for example, alluding to the user's film choice or habits such as snoring. This kind of humour seems currently difficult to implement. Overall, our findings thus imply to approach humour carefully in voice assistant design today.

\subsubsection{Always Listening Voice Assistants?}
33.3\% of dialogues were either initiated by the assistant or by the user without calling it. This lack of a wake word implies that the voice assistant was expected to always listen. In contrast, prior work suggests that users are uncomfortable with this due to privacy concerns~\cite{tabassum2019}. Braun et al.~\cite{braun2019} also reported that people have mixed opinions on whether the voice assistant should initiate conversations. One explanation for our result could be that people did not think about these implications when writing their dialogues. Nevertheless, people might also assume that their perfect voice assistant is trustworthy and thus would be more comfortable with it being allowed to listen to their conversations. 

\subsubsection{Comparison with Commercial Voice Assistants}
Comparing people's vision of a perfect voice assistant with commercially available voice assistants today (most prominently, Amazon's Alexa, Apple's Siri, and the Google Assistant), one of the most notable differences concerns the delivery of recommendations and suggestions. While most participants included suggestions and recommendations in their envisioned dialogues, today's voice assistants are designed in a way which includes recommendations only sparsely. When prompted with the scenarios used in our study, Alexa, for example, does not offer any recommendations on what to pack for a trip, while Siri only tells the weather without a specific suggestion when asked what to wear today. On the other hand, Alexa offers different suggestions for activities based on the user's current mood. Since this form of suggestion seems to be valued by people, voice assistants could offer such features more extensively. However, the envisioned dialogues suggest that personalising suggestions to individual users is likely to be challenging.

Notably, today's commercial voice assistants already implement a kind of humour similar to what people envisioned in their dialogues. For example, when asked for a good night story, Siri sarcastically replies whether the user would like a glass of warm milk next. The Google Assistant jokingly suggests overtone singing upon being asked for music recommendations. Conversely, commercial voice assistants avoid giving an opinion. For example, when asked whether the music is too loud, Siri, Alexa, and the Google Assistant turn down the volume instead of answering the question. While this seems reasonable at the moment to avoid increasing expectations~\cite{doyle2019}, future voice assistants might carefully assess whether their user enjoys humour and opinions and correspondingly decide whether to incorporate them. For example, a voice assistant could consider the current volume, time of day, the user's living situation, and past music behaviour to give an opinion.

Besides, the envisioned perfect voice assistant seems to be able to ``think'' more independently by directly presenting an answer, while commercial voice assistants often fall back on web searches. For example, when asked for movie times, most participants' envisioned the voice assistant to give an immediate answer, whereas Siri presents a web search with the results.

\subsubsection{Summary}
In summary, most people envisioned dialogues with a perfect voice assistant that were highly interactive and not purely functional; it is smart, proactive, and has personalised knowledge about the user. On the other hand, peoples' attitude towards the assistant's role and it expressing humour and opinions diverged. The envisioned characteristics echo previous findings on the need to convey voice assistant skills through dialogue~\cite{luger2016} and that few users see a voice assistant as a friend~\cite{doyle2019, clark2019}, while expanding on the importance of different user requirements for conversational skills missing at present~\cite{porcheron2018}. They challenge the assumption that users feel voice assistants should not use opinions, humour, or social talk~\cite{doyle2019} -- some users welcome this for a perfect voice assistant.

To formalise these findings, we conclude this section using the ten dimensions for conversational agent personality by~\citet{voelkel2020}: The assistant's personality envisioned here seems to be high on \textit{Serviceable}, \textit{Approachable}, \textit{Social-Inclined}, and \textit{Social-Assisting}, and low on \textit{Confrontational}, \textit{Unstable}, and \textit{Artificial}. With respect to the dimensions \textit{Social-Entertaining} and \textit{Self-Conscious}, participants seemed to have mixed opinions.

\subsection{A (Small) Effect of Personality?}
\label{sec:dis:personality}
Our exploratory analysis indicates a limited effect of personality on people's vision of a perfect voice assistant. Moreover, the significant results are to be interpreted with caution due to the number of tests performed. Figure~\ref{fig:correlations} shows correlations comparable to previous research~\cite{mehl2006, pennebaker1999}.

Our results suggest that \textit{Neuroticism} has a small negative relationship with \textit{humour}. Neurotic individuals tend to perceive new technologies as less useful and often experience negative emotions when using them since they associate them with stress~\cite{devaraj2008}. Besides, the \textit{Joke} scenario described a situation in which the user is responsible for entertaining friends -- a potentially stressful situation for a neurotic user. Therefore, neurotic individuals might prefer staying in control of the situation by telling the voice assistant exactly what to do instead of relying on its sense of humour. 

Our LMM analysis indicates \textit{Conscientiousness} as a negative predictor for the assistant offering an \textit{opinion}. When seeking information, conscientious people are described as \textit{deep divers}, valuing high quality information and structured deep analysis~\cite{heinstrom2005}. It thus seems fitting that these users prefer their assistant to provide fact-based instead of opinionated knowledge, in particular since it is difficult to assess the quality of this information.  

The correlations further suggest a small positive relation between \textit{Openness} and \textit{trusting the assistant with complex tasks}. Individuals who score high on \textit{Openness} are intellectually curious and were found to be early adopters of new technology~\cite{yoon2013}. Hence, it seems likely they might be more willing to try out new use cases. However, it could also be possible that this correlation stems from open individuals' higher creativity. 

Our findings do not indicate any meaningful relationship between \textit{Extraversion} and the characteristics of an envisioned conversation with a perfect voice assistant. This is surprising since the relationship between extraversion and linguistic features is usually most pronounced~\cite{furnham1990, oberlander2004, beukeboom2013, dewaele2000, pennebaker1999}.   

Summing up, our findings give first pointers to potential relationships between Big Five personality traits and characteristics of the envisioned dialogue with a perfect voice assistant. However, this relationship might be less pronounced than could have been expected from related work. A reason for this lack of effect could be that our work only concentrates on \textit{linguistic} content of a dialogue, while previous work particularly synthesised personality from \textit{paraverbal} features (e.g.~\cite{lee2003}). 
This opens up opportunities for future work, which we discuss in the following section.

\section{Conclusion and Future Work}
While recent work has emphasised the gulf between user expectations and voice assistant capabilities~\cite{luger2016, cowan2017, porcheron2018}, little has been known about what users actually \textit{do} want. To address this gap, we contribute a systematic empirical analysis of users' vision of a conversation with a perfect voice assistant, based on 1,835 dialogues written by 205 participants in an online study. 

Overall, our dialogues reveal a preference for human-like conversations with voice assistants, which go beyond being purely functional. In particular, they imply assistants that are smart, proactive, and include knowledge about the user. 
We further found varying user preferences for the assistant's role, as well as its expression of humour and opinions.

Since these differences between users can only be explained to a limited extent by their personality, future research should examine other user characteristics more closely to shed further light on how to make the interaction experience more personal. For example, user preference for a particular role of the voice assistant could also be due to age or current living situation. Our work also suggests that a perfect voice assistant adapts to different situations. Thus, exploring the usage context and its influence on users' vision can be another starting point for future research. Finally, our work points to the importance of a \textit{trustworthy} voice assistant that acts in the user's best interest. Given recent eavesdropping scandals about voice assistants in users' homes~\cite{washingtonpost2019}, future work should examine how this trust can be built while at the same time integrating the interests of companies.

In a wider view, our study underlines the vision of enabling \textit{conversational} UIs, rather than single command ``Q\&As''. Towards this vision, our method was effective in enabling people to depict potential experiences anchored by existing concrete use cases. Looking ahead, allowing people to draw upon their own creativity and experiences seems particularly promising in the context of user-centred design of technologies that are envisioned to permeate users' everyday lives. 

Beyond our analysis here, we release the collected dataset to the community to support further research: \textbf{\url{www.medien.ifi.lmu.de/envisioned-va-dialogues}}

\begin{acks}
We greatly thank Robin Welsch, Sven Mayer, and Ville Mäkelä for their helpful feedback on the manuscript.

This project is partly funded by the Bavarian State Ministry of Science and the Arts and coordinated by the Bavarian Research Institute for Digital Transformation (bidt), and the Science Foundation Ireland ADAPT Centre (13/RC/2106). 
\end{acks}

\bibliographystyle{ACM-Reference-Format}
\bibliography{bibliography}

\end{document}